\documentclass[aps,prl,reprint,superscriptaddress]{revtex4-2}

\usepackage{amsmath,amssymb}
\usepackage{bm}
\usepackage{graphicx}
\usepackage{hyperref}
\hypersetup{hidelinks}

\begin{document}

\title{Plasma screening and configuration interaction effects induced large enhancement on L-shell photoionization cross sections and opacity}

\author{Fuyang Zhou}
\thanks{These authors contributed equally to this work.}
\affiliation{National Key Laboratory of Computational Physics,
Institute of Applied Physics and Computational Mathematics,
Beijing 100088, P. R. China}

\author{Shengbo Niu}
\thanks{These authors contributed equally to this work.}
\affiliation{School of Optoelectronics,
University of Chinese Academy of Sciences,
Beijing 100049, P. R. China}

\author{Simei Lu}
\affiliation{School of Mathematics and Physics, Qinghai University, Xining, 810016, P. R. China}

\author{Chuangying Li}
\affiliation{National Key Laboratory of Computational Physics, Institute of Applied Physics and Computational Mathematics, Beijing, 100088, P. R. China}

\author{Xiang Gao}
\affiliation{National Key Laboratory of Computational Physics, Institute of Applied Physics and Computational Mathematics, Beijing, 100088, P. R. China}

\author{Yong Wu}
\email{wu_yong@iapcm.ac.cn}
\affiliation{National Key Laboratory of Computational Physics, Institute of Applied Physics and Computational Mathematics, Beijing, 100088, P. R. China}
\affiliation{Center for Applied Physics and Technology, HEDPS, Peking University, 100871, Beijing, P. R. China}

\author{Yizhi Qu}
\email{yzqu@ucas.ac.cn}
\affiliation{School of Optoelectronics, University of Chinese Academy of Sciences, Beijing, 100049, P. R. China}

\author{Jianguo Wang}
\affiliation{National Key Laboratory of Computational Physics, Institute of Applied Physics and Computational Mathematics, Beijing, 100088, P. R. China}

\date{\today}

\begin{abstract}
An opacity model that incorporates improved treatments of both plasma screening and configuration interaction (CI) effects is proposed, and a 25-30\% enhancement on the iron L-shell opacity is predicted at solar interior temperatures. It is originated from the plasma screening induced 14-17\% enhancement on the photoionization cross sections and the CI induced 10-20\% enhancement on photoexcitation and photoionization cross sections for open L-shell ions. These explain the long-standing discrepancy between theoretical and experimental iron opacity [Nature 517, 56], and the relatively weaker enhancements on chromium and nickel opacity [Phys. Rev. Lett. 122, 235001] due to the sensitivity of these effects to the different L-shell electron population and plasma temperature/density. This letter provides the systematic interpretation of L-shell opacity measurements at solar interior temperatures, and advances the accurate simulation of opacity and radiative transport in high-energy-density plasma.
\end{abstract}

\maketitle

\section{I. Introduction}

Radiative opacity quantifies the ability of photon absorption in matter and plays an important role on the energy transport in high-energy-density plasma. In astrophysics, opacity is crucial to the simulations of stellar structure and evolution, and its accuracy affects the precision of stellar models [1]. In recent decades, solar interior opacity has been questioned due to the discrepancy between standard solar models and helioseismic observations, which could be resolved if the theoretical mean opacity is increased by about 15\% [2–5]. Moreover, the laboratory measurements of iron ($Z=26$ ) opacity at the solar interior temperatures in the Sandia Z-pinch facility disagree considerably with predicted opacities by 30–400\% in the wavelength range of 7–13 Å [6]. More surprisingly, the chromium ($Z=24$ ) and nickel ($Z=28$ ) opacities measured with the same methods agree with the opacity models [7]. The 30\%–45\% model-data disagreements in the high-energy featureless direct photoionization (bound-free continua) was observed only from iron [6]. These systematic experimental studies reveal missing physics in opacity models.

Motivated by this problem, various theoretical efforts [8–17] have been devoted to investigate the multiple physical effects encountered in opacity models. Pain and Gilleron [8] investigated the effects of highly excited states on opacity using a partial detailed-line-accounting calculation. Krief et al. [9] demonstrated that increasing line widths by two orders of magnitude can recover the missing solar interior opacity. Fontes et al. [10] carefully analyzed relativistic effects on opacity. More et al. [11] and Pain [12] examined the contributions of two-photon and multi-photon processes to opacity. Krief et al. [13] examined the effects of ion-ion correlation on the ionic structure and opacity, and indicated that these effects indeed increase the Rosseland mean opacity near the convection zone. Zeng et al. [14] proposed that the continuum electrons lose phase and coherence in the radiative processes, giving rise to a phenomenon of transient spatial localization and considerably enhance the continuum opacity. However, the reason why the predicted opacity is lower than the data only for Fe at high temperature and density has not been clearly demonstrated. Meanwhile, these studies also reveal that modeling environmental effects of high-energy-density plasma and calculating photo absorption for open L-shell excited states involve substantial complexities, indicating fundamental limitations in approximations of previous opacity models.

In this letter, an opacity model that incorporates improved treatments of both plasma screening and configuration interaction (CI) effects is proposed to investigate this long-standing problem. This model predicts significantly enhanced photoionization cross sections due to the plasma screening effects, which is described by employing atomic-state-dependent (ASD) electron screening model [18] and ion-ion correlation (IC) model [13,19]. Additionally, the CI effect is found to be crucial on photoexcitation and photoionization cross sections for open L-shell ions. We suggest that the unexplained experimental observations and missing opacity in solar interior can be interpreted by the combined influence of these effects.

\section{II. Plasma screening effect on photoionization cross sections}

For ions embedded in the Z-pinch plasma [6,7], strong plasma screening effects will be encountered due to complicated many-body interactions with the surrounding ions and electrons, and significantly affects the atomic wave functions and the photoionization cross sections. Considering the plasma screening effect, the Dirac Hamiltonian for an atom is given by
\begin{equation}
H_{\mathrm{DC}}
=
\sum_{i=1}^{N}
\left[
c\,\boldsymbol{\alpha}_{i}\!\cdot\!\mathbf{p}_{i}
+(\beta_i-1)c^{2}
+V(r_i)
\right]
+
\sum_{i<j}^{N}\frac{1}{r_{ij}} .
\label{eq:dirac-coulomb}
\end{equation}
where $\boldsymbol{\alpha}_{i}$ and $\beta_i$ are Dirac matrices, and $\mathbf{p}_{i}$  is the momentum operator. The central atomic potential $V(r_i)$ includes the nuclear potential and the external field generated by plasma screening, and the present model represents it as
\begin{equation}
V(r)=-\frac{Z}{r}+V_{e}(r)+V_{\mathrm{ion}}(r).
\end{equation}
Here, \(r=|\mathbf{r}|\) is the distance from the center of the nucleus,
and \(Z\) is the nuclear charge. The potential
\(V_{\mathrm{e}}(r)=\int
\rho_{\mathrm{e}}(\mathbf{r}')/
|\mathbf{r}-\mathbf{r}'|\,\mathrm{d}^{3}\mathbf{r}'\)
arises from the electron--electron interaction with the plasma electrons,
whereas
\(V_{+}(r)=-\int
\rho_{+}(\mathbf{r}')/
|\mathbf{r}-\mathbf{r}'|\,\mathrm{d}^{3}\mathbf{r}'\)
is due to the penetrating positive charges associated with the surrounding
ions. Here, \(\rho_{\mathrm{e}}\) and \(\rho_{+}\) denote the spatial
distributions of the plasma electrons and positive charges, respectively.
The ASD model[18] is employed to calculate the plasma-electron
distribution \(\rho_{\mathrm{e}}\) and the resulting screening effect.
Within a unified framework, the model accounts for contributions from
both elastic and inelastic collisions between the ion and plasma electrons.
Its feasibility and validity have been demonstrated by successfully
reproducing recent experimental results for spectral line shifts and
ionization-potential depression (IPD) over wide ranges of temperature
and density[18,20].

\begin{figure*}[t]
    \centering
    \includegraphics[width=1.8\columnwidth]{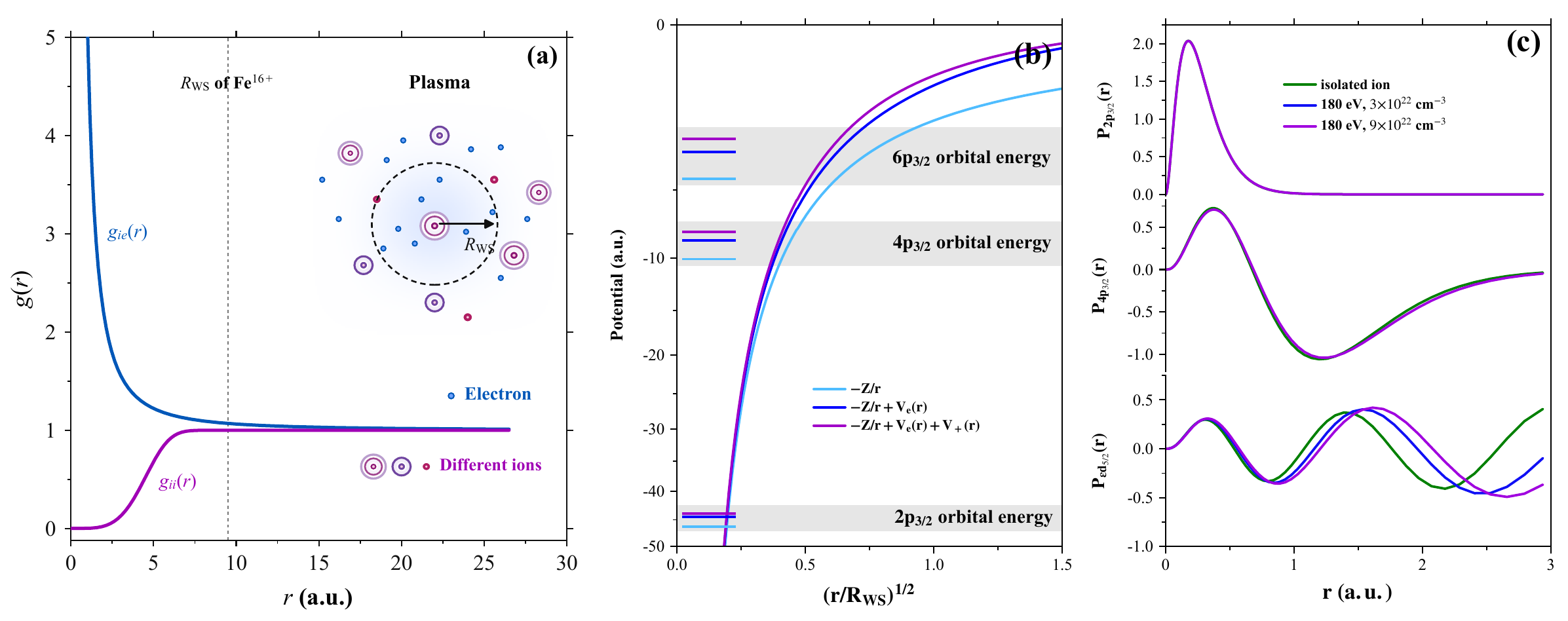}
    \caption{
        Plasma screening effects on atomic radial wave functions. (a) The pair distribution functions of ion-electron and ion-ion of $Fe^{16+}$ ion embedded in plasma with temperature $T_e = 180eV$  and mean electron number density $\rho_0 = 3\times 10^{22} \text{cm}^{-3}$. $R_WS = \sqrt{3Z^*/(4\pi \rho_0)}$ is the Wigner-Seitz radius of $Fe^{16+}$. (b) The central atomic potentials $V(r)$  in the present model, as well as the energies of $2p_{3/2}$, $4p_{3/2}$ and $6p_{3/2}$ orbitals calculated with these potentials. (c) Taking $np \to \epsilon d$ photoionization as an examples, the plasma screening effects on the atomic radial wave functions of bound electron orbital ($2p_{3/2}$, $4p_{3/2}$)  and continuum orbital ($\epsilon p_{3/2}$  with $\epsilon =0.1 a.u.$) are presented.
    }
    \label{fig1}
\end{figure*}

As shown in Figure 1(a) for $Fe^{16+}$ ion embedded in a dense plasma, the ion-electron and ion-ion pair distribution functions indicate that the surrounding ions and electrons of hot dense plasma can approach the nucleus via collision processes and thereby screen the Coulomb interaction between nucleus and bound electron. From figure 1(b), it can be found from that the screening effects from surrounding ions become significant for excited orbitals. But the commonly used ion-sphere model, which assumes each ion is enclosed within a sphere of radius $R_{WS}$, underestimated its impact [13,19]. Therefore, to accurately describe the screening effect on excited atomic orbitals, we extend the ASD model by incorporating the screening from surrounding ions within the IC model [13,19], which allows ionic charges to penetrate the ion sphere and contributes to screening of the central potential. In this model, charge neutrality requires that $\rho_{e}(r)\rightarrow\rho_{+}(r)\rightarrow\rho_{0}$  as $r \to \infty$ , and
\begin{equation}
\rho_0=\sum Z^*N_{ion}^i,
\end{equation}
where $Z^*$ is the nuclear charge minus the number of bound electrons, $N_{ion}^i$ is the ion density of ion “i”. For the IC model, we have
\begin{equation}
\rho_+(r)=\sum_iZ^*\rho_0g_{ii}(r)
\end{equation}

\begin{equation}
\mathrm{and}\int_0^\infty[\rho_b(r)+\rho_e(r)-\rho_+(r)]d^3r=Z.
\end{equation}
$\rho_b(r)$  is the distribution of bound electrons, $g_{ii}(r)$ is the ion-ion pair distribution function and it has to satisfy the hypernetted-chain (HNC) integral equation. The distributions $\rho_e(r)$  and $\rho_+(r)$ are obtained through a self-consistent iterative procedure.

Taking the $Fe^{16+}$ ion embedded in plasma with $T_e = 180eV$ and $\rho_0  = 3\times10^{22} \text{cm}^{-3}$ as an example, figure 1(b) and 1(c) show that the change of central atomic potential $V(r)$  from plasma screening effect significantly affects the atomic energy levels and wave functions of continuum electrons, resulting in ionization potential depression (IPD) and remarkable changes in the photoionization cross sections. As the density $\rho_0$  increases to $\rho_0  = 9\times10^{22} \text{cm}^{-3}$ , the screening effect further intensifies its modification of both bound and continuum wave functions. Figure 1(c) reveals that the changes in the central atomic potential $V(r)$  significantly broaden the spatial range of continuum orbitals and enhances the spatial overlap between bound and continuum wave functions, ultimately leading to an increase of the photoionization cross sections.

\begin{figure}[t]
    \centering
    \includegraphics[width=\columnwidth]{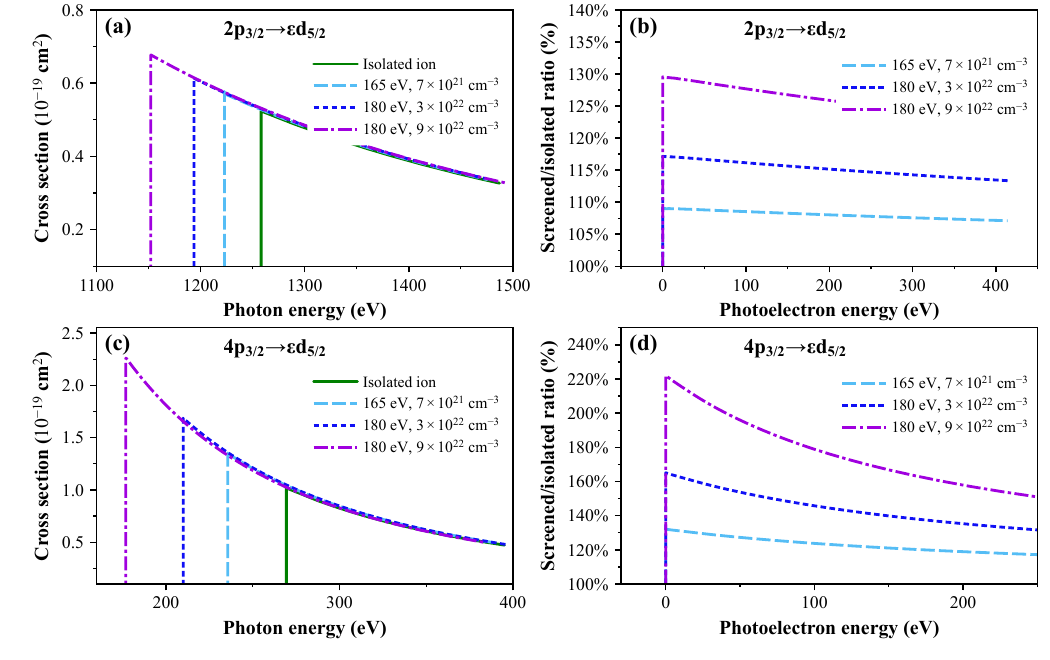}
    \caption{
        Plasma screening effects on photoionization cross sections in dense
        plasma. (a) Photoionization cross sections for the
        \(2p_{3/2}\to\varepsilon d_{5/2}\) transition channel of
        Fe\(^{16+}\) embedded in different plasma environments as a function
        of photon energy. (b) Ratios of the screened cross sections to the
        isolated-ion results for the
        \(2p_{3/2}\to\varepsilon d_{5/2}\) transition channel as a function
        of photoelectron energy. (c) Photoionization cross sections for the
        \(4p_{3/2}\to\varepsilon d_{5/2}\) transition channel.
        (d) Corresponding cross-section ratios for the
        \(4p_{3/2}\to\varepsilon d_{5/2}\) transition channel.
    }
    \label{fig2}
\end{figure}

Figures~2(a) and 2(b) present the photoionization cross sections for the
\(2p_{3/2}\rightarrow\varepsilon d_{5/2}\) transition channel of
Fe\(^{16+}\) embedded in different plasma environments. Plasma screening
leads to ionization-potential depression (IPD) and significantly enhances
the photoionization cross sections by increasing the spatial overlap
between the bound and continuum atomic wave functions. Moreover, this
enhancement becomes stronger as the electron density increases.
At \(T_{\mathrm{e}}=180\,\mathrm{eV}\) and
\(\rho_{0}=7.1\times10^{21}\,\mathrm{cm}^{-3}\), the photoionization cross
section for the \(2p_{3/2}\rightarrow\varepsilon d_{5/2}\) channel is
predicted to increase by \(7\)--\(9\%\). The corresponding enhancements
reach \(14\)--\(17\%\) at
\(\rho_{0}=3.0\times10^{22}\,\mathrm{cm}^{-3}\), corresponding to the
conditions achieved at the Sandia \(Z\)-pinch facility, and
\(23\)--\(29\%\) at
\(\rho_{0}=9.0\times10^{22}\,\mathrm{cm}^{-3}\), corresponding to
conditions near the boundary between the solar convection and radiative
zones.
Furthermore, because excited orbitals have a greater radial extent, their
photoionization processes are more strongly affected by plasma screening.
As shown in Figs.~2(c) and 2(d), the photoionization cross section for the
\(4p_{3/2}\rightarrow\varepsilon d_{5/2}\) transition channel exhibits a
larger enhancement than that for the
\(2p_{3/2}\rightarrow\varepsilon d_{5/2}\) channel.

\section{III. Configuration interaction effect on photoionization and opacity}

At solar interior temperatures, iron is ionized to charge states of 16+ and higher, generating numerous complex open L-shell excited states. Consequently, it is essential to examine the CI effect on the L-shell photoionization and opacity of these Fe ions. In particular, mainstream opacity codes typically adopt one orbital basis, which is usually obtained from Dirac-Hartree-Fock calculations on the ground configuration, to compute all configurations. Such a basis is not fully appropriate for excited configurations, and the CI treatment may be inadequate. To address this, we employ the Flexible Atomic Code (FAC) [21] to examine the CI effects on the L-shell photoionization of both closed-shell ground configurations and open-shell excited configurations. As an example, figure 3 illustrates the photoionization cross sections obtained with CI and single-configuration (SC) for closed L-shell state $2s^2 2p^6 (^1 S)$ and open L-shell state $2s^2 2p^5 3p (^1 S_0)$ of isolated $Fe^{16+}$. For the closed L-shell state $2s^2 2p^6$, the direct photoionization cross sections calculated using SC and CI methods are identical, and difference arise only in the indirect resonance photoionization processes, which proceed via photoexcitation into autoionizing states. Due to the CI effects, the increase in resonance ionization channels leads to local enhancements to the photoionization cross sections, as demonstrate by the additional resonance peak observed around 1650 eV in the CI calculations of figure 3(a). But for the open L-shell state $2s^2 2p^5 3p$, the CI effect can influences the direct photoionization cross section by approximately 10\%, as shown in figure 3(b). Additionally, for indirect resonance photoionization, the CI effect modifies both resonance positions and intensities, leading to an increased photoionization cross section at higher photon energies. Subsequently, large-scale CI calculations for a broad range of atomic parameters are carried out to investigate the opacity at solar interior temperatures.

\begin{figure}[t]
    \centering
    \includegraphics[width=\columnwidth]{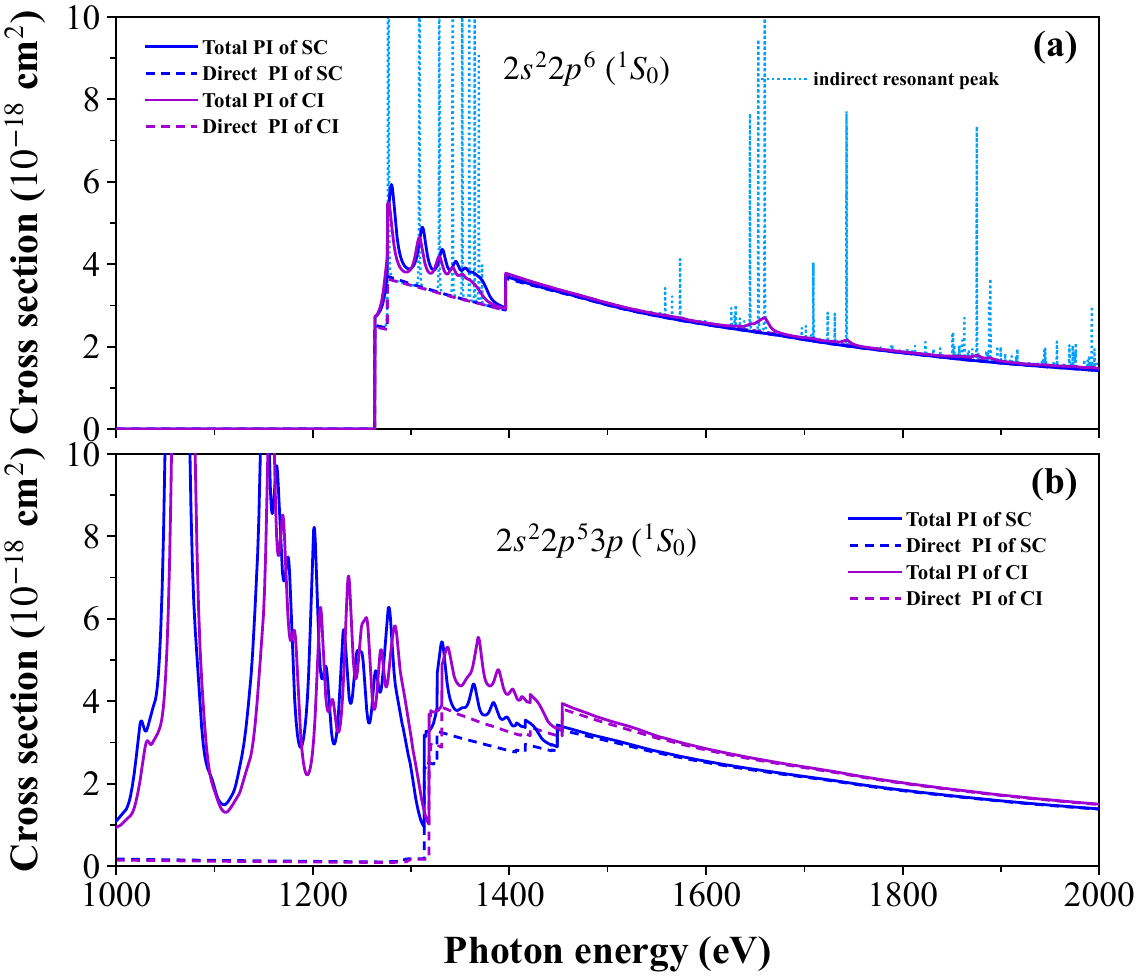}
    \caption{
        Configuration-interaction (CI) effects on the photoionization
        (PI) cross sections of the ground state
        \(2s^{2}2p^{6}\,({}^{1}S_{0})\) and the excited state
        \(2s^{2}2p^{5}3p\,({}^{1}S_{0})\) of Fe\(^{16+}\).
        The CI calculations include configurations generated by single
        and double excitations from the valence orbitals to orbitals
        with principal quantum number \(n\leq 8\).
        Indirect resonant photoionization processes are also included
        using the isolated-resonance approximation.
        To illustrate their effects on the opacity, the widths of all
        resonance peaks are uniformly set to \(10\,\mathrm{eV}\).
    }
    \label{fig:ci-photoionization}
\end{figure}

Figure 4(a) and 4(b) illustrate the CI effect on the bound-bound (BB) and bound-free (BF) opacities of $Fe^{15+}$ and $Fe^{16+}$ ions at $T_e=180 eV$. Here, BB opacity includes contributions from all bound-state photoexcitation and resonance photoionization, while BF opacity arises from direct photoionization. For $Fe^{15+}$ ion in figure 4(a), where all bound states have a closed L-shell, the CI effect on the opacity contributed from both L-shell BB transitions and photoionizations is almost negligible. However, for $Fe^{16+}$ ion in figure 4(b), CI significantly alters the calculated results for photoexcitation and photoionization from open L-shell excited states, resulting in an approximate 10-20\% increase in both BB and BF opacities within photon energy of 1250-1600 eV. Overall, the CI effect is found to be significant in opacity calculations for Fe ions with open L-shell. 

\begin{figure}[t]
    \centering
    \includegraphics[width=\columnwidth]{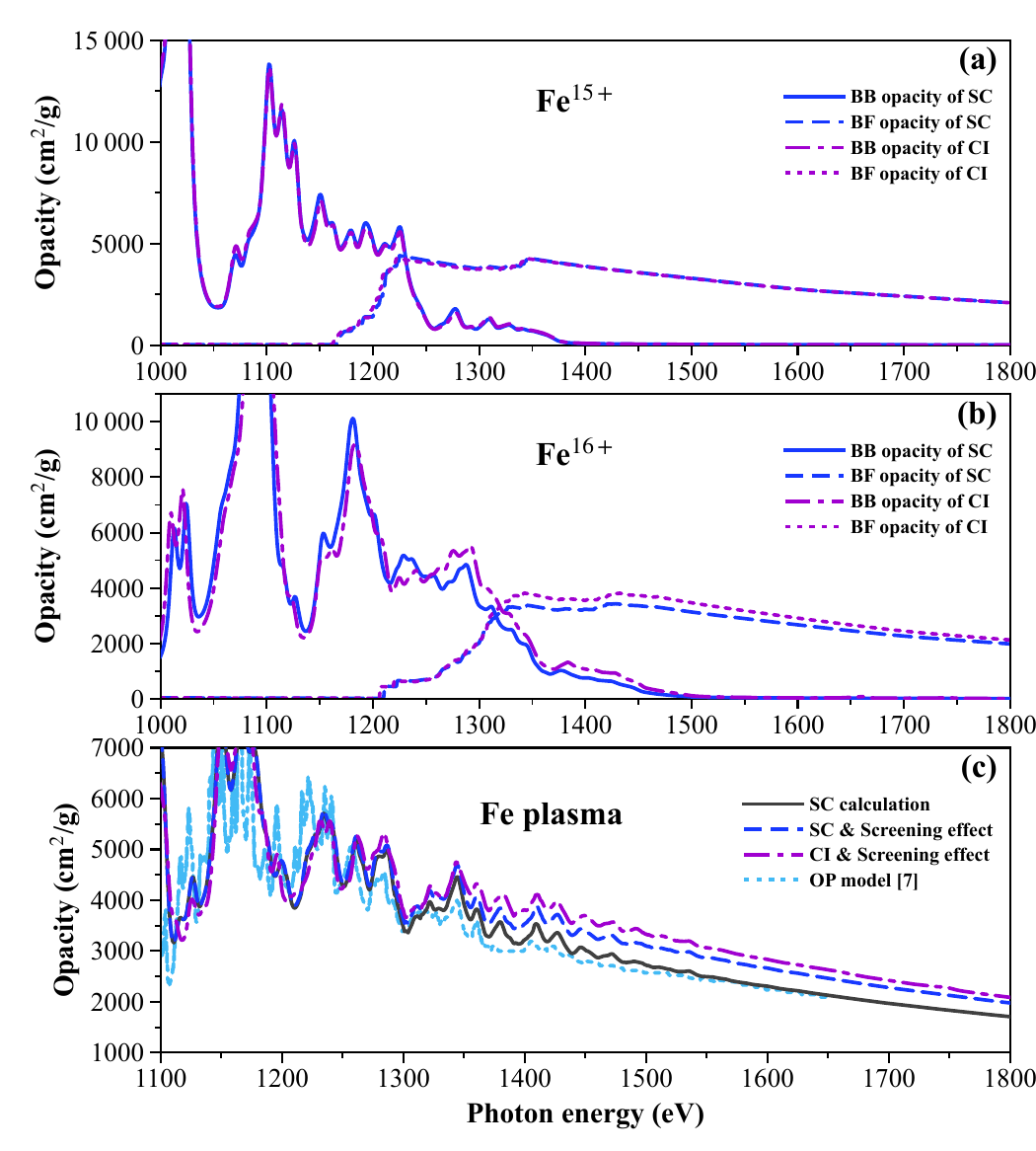}
    \caption{
        Configuration interaction effects on iron opacity. Configuration interaction effects onon BB and BF opacities of $Fe^{15+}$ and $Fe^{16+}$ ion at $T_e=180 eV$ are presented in plots (a) and (b). In the CI calculations, all energy levels, transition rate, and photoionization cross sections were calculated based on FAC code with considering the configuration obtained from single excitation from both initial and final contributions to orbitals $n\leq 6$. (c) Configuration interaction effects on total opacity of Fe plasma at $T_e = 180eV$ and $n_e = 3.1 \times 10^{22} \text{cm}^{-3}$.
    }
    \label{fig1a}
\end{figure}

As shown in the figure 4(c), for total opacity of Fe plasma at $T_e =180$  and $n_e=3.1\times 10^{22} \text{cm}^{-3}$, our SC calculation are generally consistent with the OP results [7]. Plasma screening and CI effects increase the opacity for photon energies above 1250 eV by about 17\% and 8\%, respectively. As a result, the refined calculation results are 25-30\% larger than those from the SC and OP models. The results reveal that incomplete consideration of these CI and plasma screening effects may lead to theory-experiment discrepancies exceeding the measurement uncertainty. It is noted that since plasma screening effects are applied only to the direct photoionization cross sections in this work, the coupling between screening and CI effects is not fully included and requires self-consistent treatment in future studies.

\section{IV. Analyses and explanation of opacity experiments on the Z-pinch}

After considering the plasma screening and CI effects discussed above, the calculated spectrally resolved opacity of Cr, Fe and Ni plasmas are compared with the experiments on the Z-pinch [6,7] and theoretical results of OP [7] in figure 5. The present theory is in much closer agreement with the experiment, and there are two main improvements of our new results. Firstly, due to the approximately 25-30\% enhancement of opacity from both plasma screening and CI effect, the present work well explains the experimental opacity within photon energies of 1200-1500 eV for Fe plasma at $182eV$ and $3.1\times 10^{22} \text{cm}^{-3}$, as shown in figure 5(b). It is noted that the experiments actually measured the transmissivity of FeMg mixture. After including the screening effect on the transmissivity of Mg, the original Fe calculations shown by the dashed blue and magenta curves in figure 5(b) are corrected to the corresponding solid curves. The correction is most visible above 1500 eV, resulting in good agreement between the theory and experiment for the entire spectrally resolved opacity spectrum. The details of the correction are provided in supplemental material. Secondly, as seen in Figures 5(a) and 5(c), the measured opacities of Cr and Ni at similar temperature and density are also reasonably reproduced by the present model. Due to the sensitivity of CI and plasma screening effects to the L-shell electron population and plasma temperature/density, the corresponding enhancements on Cr and Ni opacity are relatively weaker, e.g. about 15-18\% enhancement for Cr. In general, the opacities calculated for Cr, Fe, and Ni plasmas agree with the measure data within the experimental uncertainties. The remaining difference may arise from the uncertainties in temperature and density, as well as the non-equilibrium effects on the populations of ions and their excited states. By appropriately varying temperature and density, all calculated results can be brought into better agreement with experimental values.

\begin{figure}[t]
    \centering
    \includegraphics[width=\columnwidth]{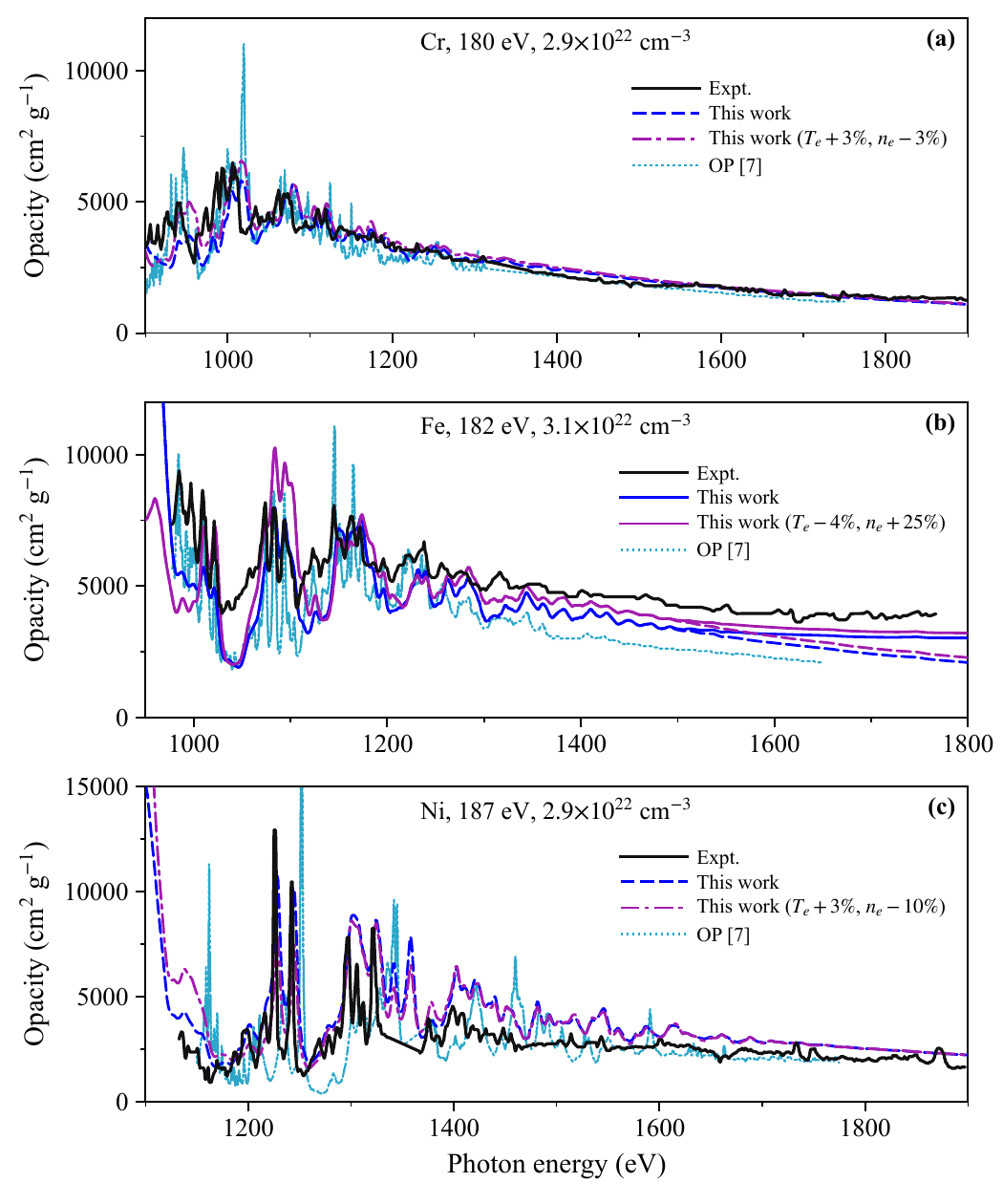}
    \caption{
        Comparison of the calculated spectrally resolved opacities of Cr,
        Fe, and Ni plasmas with the experimental results of
        Refs.[6,7] and the OP calculations of
        Ref.[7]: (a) Cr plasma at
        \(T=180\,\mathrm{eV}\) and
        \(n_{\mathrm{e}}=2.9\times10^{22}\,\mathrm{cm}^{-3}\);
        (b) Fe plasma at
        \(T=182\,\mathrm{eV}\) and
        \(n_{\mathrm{e}}=3.1\times10^{22}\,\mathrm{cm}^{-3}\);
        and (c) Ni plasma at
        \(T=187\,\mathrm{eV}\) and
        \(n_{\mathrm{e}}=2.9\times10^{22}\,\mathrm{cm}^{-3}\).
        Based on the temperature and density uncertainties from refs. [6,7], the impacts of these uncertainties on opacity calculations are also shown. In panel (b), the dashed blue and magenta curves show the original Fe calculations, whereas the corresponding solid curves include the Mg correction described in the supplemental material.
    }
    \label{fig:opacity-comparison}
\end{figure}

\section{V. Conclusion}

Opacity calculations involve challenging issues in atomic and plasma physics, especially in accounting for complex quantum many-body effects such as plasma screening and configuration interaction. By employing the ASD and IC models, the screening effects from both electrons and ions of plasma on the photoionization cross sections are investigated and found to be significant for dense plasma. Additionally, the CI effect is found to play an important role in opacity calculations of open L-shell ions. By incorporating the improved treatments of both plasma screening and CI effects, an novel opacity model is proposed and provides a reasonable interpretation of the systematic experimental opacity data obtained at solar interior temperatures from the Sandia Z-pinch facility. This work advances the understanding of radiative opacity of high-energy-density plasma, and indicates that the missing opacity in solar interior can be interpreted by precisely treatments of plasma screening and CI effects.

\begin{acknowledgments}
This work was supported by the National Natural Science Foundation of China (Grants No. U2430208). We also acknowledge the National Key Laboratory of Computational Physics.
\end{acknowledgments}

\end{document}


\begin{center}
  {\Large\bfseries Supplemental material}
\end{center}

\section*{I. Effect of the Mg subtraction on the inferred Fe opacity}

The Fe opacity reported by Bailey et al.~\cite{Bailey2015} was inferred from
transmission measurements through mixed Fe/Mg samples, the measured optical
depth of which is
\begin{equation}
  \tau_{\mathrm{Fe+Mg}}^{\mathrm{meas}}(E)
  =
  \Sigma_{\mathrm{Fe}}\kappa_{\mathrm{Fe}}(E)
  +
  \Sigma_{\mathrm{Mg}}\kappa_{\mathrm{Mg}}(E).
  \label{eq:S1}
\end{equation}
Here, \(E\) is the photon energy,
\(\tau_{\mathrm{Fe+Mg}}^{\mathrm{meas}}\) is the measured optical depth of the
mixed Fe/Mg sample, \(\Sigma_{\mathrm{Fe}}\) and
\(\Sigma_{\mathrm{Mg}}\) are the Fe and Mg areal mass densities, and
\(\kappa_{\mathrm{Fe}}\) and \(\kappa_{\mathrm{Mg}}\) are the Fe and Mg
opacities. The Fe opacity inferred from the mixed sample is given by
\begin{equation}
  \kappa_{\mathrm{Fe}}^{\mathrm{inf}}(E)
  =
  \frac{
    \tau_{\mathrm{Fe+Mg}}^{\mathrm{meas}}(E)
    -
    \Sigma_{\mathrm{Mg}}\kappa_{\mathrm{Mg}}^{\mathrm{ref}}(E)
  }{\Sigma_{\mathrm{Fe}}},
  \label{eq:S2}
\end{equation}
where \(\kappa_{\mathrm{Fe}}^{\mathrm{inf}}\) is the inferred Fe opacity and
\(\kappa_{\mathrm{Mg}}^{\mathrm{ref}}\) is the calculated Mg opacity from the
PrismSPECT model used in Ref.~\cite{Bailey2015}. If the calculated Mg opacity
is underestimated, a residual Mg contribution persists in the inferred Fe
opacity derived from the experimental work~\cite{Bailey2015}. Since plasma
screening significantly affects the calculated Fe opacity and transmission as
discussed in the main text, the corresponding effect on the Mg opacity must
also be further examined.

\setcounter{figure}{1}
\begin{figure*}[htbp]
  \centering
  \includegraphics[width=0.82\textwidth]{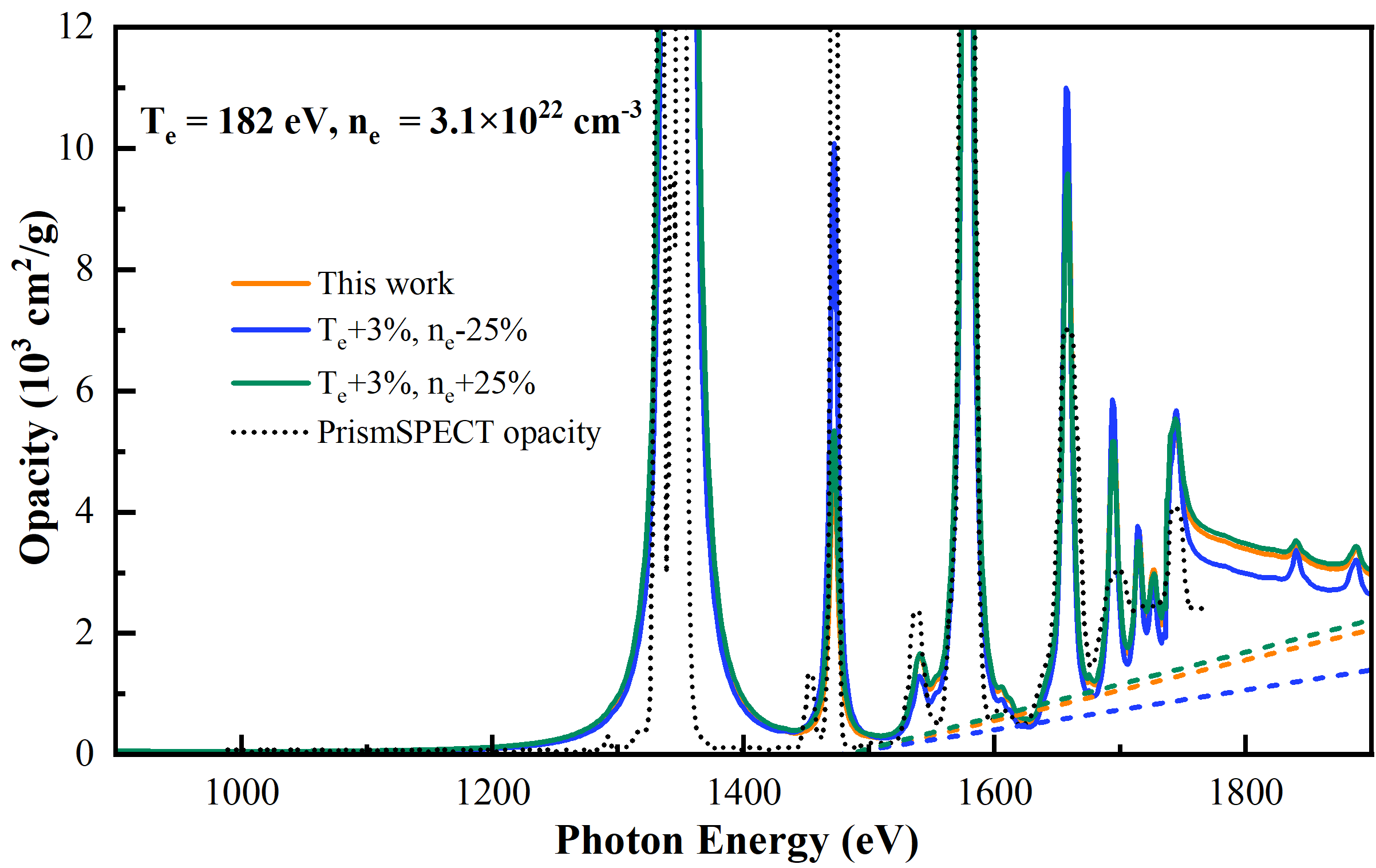}
  \caption{Comparison of Mg opacity obtained with different methods. The
  orange solid curve is the present Mg opacity at
  \(T_e=182~\mathrm{eV}\) and
  \(n_e=3.1\times10^{22}~\mathrm{cm}^{-3}\). The blue and green solid curves
  show the impact of changing plasma conditions, with \(T_e\) increased by
  \(3\%\) and \(n_e\) varied by \(-25\%\) and \(+25\%\). The black dotted
  curve is the PrismSPECT Mg opacity from Ref.~\cite{Bailey2015}. The colored
  dashed lines represent the differences in Mg opacity between the present
  calculations and the PrismSPECT results, which are approximately linear
  corrections.}
  \label{fig:S2}
\end{figure*}

Figure~\ref{fig:S2} compares the Mg opacity used for the subtraction in
Ref.~\cite{Bailey2015} with our Mg calculations including configuration
interaction and plasma-screening effects. In the bound-free region, the
present Mg opacity is found to be systematically higher than the PrismSPECT Mg
opacity due to the plasma-screening effect. In order to reduce the interference
from resonance peaks, we approximate the Mg opacity difference
\(\Delta\kappa_{\mathrm{Mg}}(E)
=\kappa_{\mathrm{Mg}}^{\mathrm{present}}(E)
-\kappa_{\mathrm{Mg}}^{\mathrm{ref}}(E)\)
using a linear interpolation between its values in the bound-free region,
taken at \(1506\) and \(1760~\mathrm{eV}\). Here,
\(\Delta\kappa_{\mathrm{Mg}}^{\mathrm{lin}}\) represents the linear
interpolation of the Mg opacity difference and is given by
\begin{equation}
  \Delta\kappa_{\mathrm{Mg}}^{\mathrm{lin}}(E)
  =
  \Delta\kappa_{\mathrm{Mg}}(1506~\mathrm{eV})
  +
  \frac{
    \Delta\kappa_{\mathrm{Mg}}(1760~\mathrm{eV})
    -
    \Delta\kappa_{\mathrm{Mg}}(1506~\mathrm{eV})
  }{1760-1506}
  (E-1506).
  \label{eq:S3}
\end{equation}
From Eq.~\eqref{eq:S2}, the above Mg correction to the Fe opacity is defined
as
\begin{equation}
  \Delta\kappa_{\mathrm{Fe}}^{\mathrm{corr}}(E)
  =
  \eta\,\Delta\kappa_{\mathrm{Mg}}^{\mathrm{lin}}(E),
  \qquad
  \eta
  \equiv
  \frac{\Sigma_{\mathrm{Mg}}}{\Sigma_{\mathrm{Fe}}}.
  \label{eq:S4}
\end{equation}
Here, \(\Sigma_{\mathrm{Mg}}\) and \(\Sigma_{\mathrm{Fe}}\) are the areal mass
densities of Mg and Fe in the mixed sample, respectively.

For comparison, we use the atomic areal densities for Z2624, as given in
Extended Data Fig.~2d of Ref.~\cite{Bailey2015}, which are
\(N_{\mathrm{Fe}}=0.926\times10^{18}~\mathrm{cm}^{-2}\) and
\(N_{\mathrm{Mg}}=1.40\times10^{18}~\mathrm{cm}^{-2}\). The ratio of areal
mass densities is given by
\begin{equation}
  \eta
  =
  \frac{24.30\,N_{\mathrm{Mg}}}{55.85\,N_{\mathrm{Fe}}}
  =
  0.658.
  \label{eq:S5}
\end{equation}
Overall, the experimental Fe opacity inferred in Ref.~\cite{Bailey2015} should
be compared with the calculated results that include the estimated residual Mg
contribution:
\begin{equation}
  \kappa_{\mathrm{Fe}}^{\mathrm{calc}}(E)
  +
  \Delta\kappa_{\mathrm{Fe}}^{\mathrm{corr}}(E)
  =
  \kappa_{\mathrm{Fe}}^{\mathrm{calc}}(E)
  +
  \eta\,\Delta\kappa_{\mathrm{Mg}}^{\mathrm{lin}}(E).
  \label{eq:S6}
\end{equation}

\begin{figure*}[htbp]
  \centering
  \includegraphics[width=0.88\textwidth]{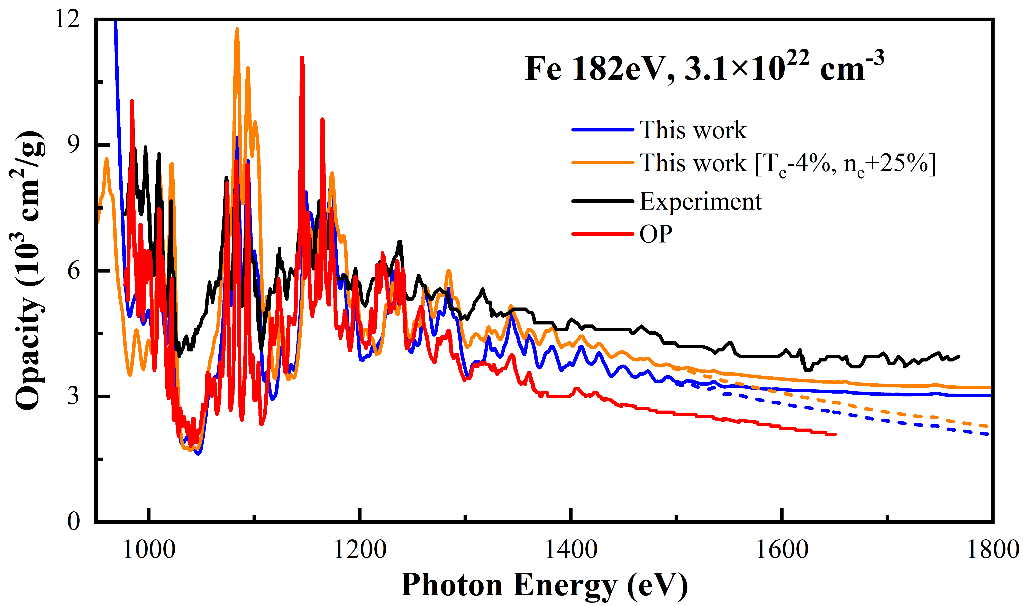}
  \caption{Effect of the estimated residual Mg contribution on the Fe opacity.
  Dashed curves show the original Fe calculations, and solid curves are
  obtained by adding the areal-density-scaled residual Mg contribution. The
  black curve is the experimental Fe opacity inferred from
  Ref.~\cite{Bailey2015}, and the red curve is the OP
  opacity~\cite{Nagayama2019}.}
  \label{fig:S3}
\end{figure*}

Figure~\ref{fig:S3} shows the effect of the estimated residual Mg contribution
on the Fe opacity. The dashed curves are the original Fe opacity calculations,
whereas the solid curves include the correction
\(\Delta\kappa_{\mathrm{Fe}}^{\mathrm{corr}}(E)
=\eta\,\Delta\kappa_{\mathrm{Mg}}^{\mathrm{lin}}(E)\). The correction
significantly affects the opacity in the high-energy bound-free region, where
the Mg opacity used in the original subtraction is lower than that from the
present calculation.

For the Cr and Ni opacity measurements, the corresponding Mg areal densities
are unavailable, and the theoretical Mg opacity used in the experimental
subtraction was not reported. Moreover, the Cr and Ni opacity data were
obtained primarily at sample areal densities of approximately
\(3\times10^{18}~\mathrm{cm}^{-2}\)~\cite{Nagayama2019}, roughly three times
higher than the Fe areal density discussed above. Since the Mg-subtraction
error on the inferred opacity of element X scales as
\(\Sigma_{\mathrm{Mg}}/\Sigma_{\mathrm{X}}\), the associated uncertainty
should be significantly smaller for Cr and Ni than for Fe. Therefore, no
similar correction was applied to the Cr and Ni opacities presented in the
main text.